PART 2

# Discourses and Myths About AI

CHAPTER 6

# The Language Labyrinth: Constructive Critique on the Terminology Used in the AI Discourse

Rainer Rehak

## Introduction

In the seventies of the last century, the British physicist and science fiction writer Arthur C. Clarke coined the phrase of any sufficiently advanced technology being indistinguishable from magic – understood here as mystical forces not accessible to reason or science. In his stories Clarke often described technical artefacts such as anti-gravity engines, 'flowing' roads or tiny atom-constructing machinery. In some of his stories, nobody knows exactly how those technical objects work or how they have been constructed, they just use them and are happy doing so.

In today's specialised society with a division of labour, most people also do not understand most of the technology they use. However, this is not a serious problem, since for each technology there are specialists who understand, analyse and improve the products in their field of work – unlike in Clarke's worlds. But since they are experts in few areas and human lifetime is limited, they are, of course, laypersons or maybe hobbyists in all other areas of technology.

After the first operational universal programmable digital computer – the Z3 – had been invented and built in 1941 in Berlin by Konrad Zuse, the rise of the digital computer towards today's omnipresence started. In the 1960s, banks, insurances and large administrations began to use computers, police and intelligence agencies followed in the 1970s. Personal computers appeared and around that time newspapers wrote about the upcoming 'electronic revolution' in publishing. In the 1980s professional text work started to become digital and in the 1990s the internet was opened to the general public and to commercialisation. The phone system became digital, mobile internet became available and in the mid-2000s smartphones started to spread across the globe (Passig and Scholz 2015).

During the advent of computers, they were solely operated by experts and used for specialised tasks such as batch calculations and book-keeping at large scale. Becoming smaller, cheaper, easier to use and more powerful over time, more and more use cases emerged up to the present situation of computer ubiquity. More applications, however, also meant more impact on personal lives, commercial activities and even societal change (Coy 1992). The broader and deeper the effects of widespread use of networked digital computers became, the more pressing political decisions about their development and regulation became as well.

The situation today is characterised by non-experts constantly using computers, sometimes not even aware of it, and non-experts making decisions about computer use in business, society and politics – from schools to solar power, from cryptography to cars. The only way to discuss highly complex computer systems and their implications is by analogies, simplifications and metaphors. However, condensing complex topics into understandable, discussable and then decidable bits is difficult in at least two ways. First, one has to deeply understand the subject and second, one has to understand its role and context in the discussion to focus on the relevant aspects. The first difficulty is to do with knowledge and lies in the classical technical expertise of specialists. But the second difficulty concerns what exactly should be explained in what way. Depending on the context of the discussion, certain aspects of the matter have to be explicated using explanations, metaphors and analogies highlighting the relevant technical characteristics and implications. Seen in this light, this problem of metaphors for technology is not only philosophically highly interesting but also politically very relevant. Information technology systems are not used because of their actual technical properties, but because of their assumed functionality, whereas the discussion about the functionality is usually part of the political discourse itself (Morozov 2013).

Given the complexity of current technology, only experts can understand such systems, yet only a small number of them actively and publicly take part in corrective political exchanges about technology. Especially in the field of artificial intelligence (AI) a wild jungle of problematic terms is in use. However, as long as discussions take place among AI specialists those terms function just as domain-specific technical vocabulary and no harm is done. But domain-specific language often diffuses into other fields and then easily loses its context, its specificity and its limitations. In this process terms which might have started as pragmatic 'weak' metaphors within the technical field, then develop into proper technical terms eventually starting to be seen as 'proper' metaphors outside their original professional context. In addition to the effect of specific terms, those metaphors can also unfold effects beyond concrete technologies but also fuel or inhibit larger narratives around them or digital technology in general. Hence, powerful metaphors push the myths of unlimited potential of (computer) technology, the superiority of computation over human reasoning (Weizenbaum 1976) or the leading role of the 'digital sublime' in transforming society (Mosco 2004). On the other hand, less colourful and less visionary metaphors keep such myths at bay and narratives grounded. Of course, it would be short-sighted to interpret the choice, development and dissemination of technical metaphors and specifically AI terminologies purely as a somehow chaotic process of misunderstandings and unclear technical usage. Those discourses, as all discourses, are a playing field of interests and power where actors brawl over the 'proper' narration either because they find sincere truth in it (e.g., transhumanist zealots of the singularity) or because it plainly benefits them politically or financially (e.g., companies selling AI), or both. Practically speaking, if relevant decision-makers are convinced that AI can develop a real 'understanding' or properly 'interpret' issues, its regular use for sensitive tasks like deciding about social benefits, guiding education, measuring behavioural compliance or judging court cases problematically looms; and corresponding companies will then eagerly come forward to sell such systems to them.

All the above dynamics motivate this work to scrutinise the AI discourse regarding its language and specifically its metaphors. The paper analyses central notions of the AI debate, highlights their problematic consequences and contributes to the debate by proposing more fitting terminology and hereby enabling more fruitful debates.

*Conceptual Domains and Everyday Language*

Unlike the abstract field of mathematics, where most technical terms are easily spotted as such, AI makes heavy use of anthropomorphisms. Considering AI-terms such as 'recognition', 'learning', 'acting', 'deciding', 'remembering', 'understanding' or even 'intelligence' itself, problems clearly loom across all possible conversations. Of course, many other sciences also use scientific terms that are derived from everyday language. In this case, these terms then have clearly defined meanings or at least linked discourses reflecting upon them. Examples are the terms 'fear' in psychology, 'impulse' in physics, 'will' in philosophy or 'rejection' in geology and 'ideology' mathematics. Often the same words have completely different meanings in different domains, sometimes even contradictory meanings, as the examples of 'work' in physics and economic theory (energy transfer via application of a force while moving an object vs. planned and purposeful activity of a person to produce goods or services) or 'transparency' in computer science and political science (invisibility vs. visibility) illustrate.

Hence, problems arise when these scientific terms are transferred carelessly into other domains or back into everyday language used in political or public debates. This can occur through unprofessional science journalism, deliberate inaccuracy for PR purposes, exaggerations for raising third-party funding, or generally due to a lack of sensitivity to the various levels and contexts of metaphors.

## The Case of Artificial Intelligence

For some years now, technical solutions utilising artificial intelligence are widely seen as means to tackle many fundamental problems of mankind. From fighting the climate crisis, tackling the problems of ageing societies, reducing global poverty, stopping terror, detecting copyright infringements or curing cancer to improving evidence-based politics, improving predictive police work, local transportation, self-driving cars and even waste removal.

### *Definitions*

The first step towards a meaningful discussion about AI would be to define what exactly one means when talking about AI. Historically there have been two major understandings of AI: strong AI or Artificial General Intelligence (AGI) and weak AI or Artificial Narrow Intelligence (ANI). The goal of AGI is the creation of an artificial human like intelligence, so an AI system with true human-like intelligence including perception, agency, consciousness, intentions and maybe even emotions (see Turing 1950 or more popular Kurzweil 2005). ANI, on the other hand, refers to very domain-specific AI systems being able to accomplish very specific tasks in very narrowly defined contexts only. Questions of agency or consciousness do not arise with ANI systems, they are merely tools, although potentially very powerful tools.

So far and tellingly, AGI can only be found in manifold media products within the fantasy or science fiction genre. Famous examples are *Samantha* in 'HER', *Data* in 'Star Trek', *HAL 9000* in '2001' (based on a novel of the aforementioned writer Arthur C. Clarke), *Bishop* in 'Aliens', the *Terminator* in the movie series of the same name or even the *Maschinenmensch* in 'Metropolis' (Hermann 2020).

In contrast, ANI systems are the ones calculating the moves in advanced chess games, the ones enhancing smartphone pictures, the ones doing pattern recognition concerning speech (e.g., natural language processing) or images (e.g., computer vision) or even the ones optimising online search results. Furthermore, within the ANI discourse mainly two more specific definitions should be mentioned. The first one focuses on the technical process of how ANI works and goes along the lines of AI being computer algorithms that improve automatically through experience (cf. Mitchell 1997 about machine learning). The second one focuses more on the phenomenon of ANI by defining AI broadly as computer systems that are able to perform tasks normally requiring human intelligence (Gevarter 1985).

Technically there are a multitude of approaches to actually build AI systems. Those approaches are usually referred to as the field of machine learning (ML) and comprise the so-called symbolic approaches with explicit data representations of relevant information like simple decision trees or formal logic-based ones like knowledge databases with inference algorithms. These approaches are comparatively limited due to the necessity of explicit data representation. Then again there are the more recent sub-symbolic approaches of ML which do not use explicit data representations of relevant information but mathematical (e.g., statistical) methods for processing all kinds of data. Artificial neural networks (ANN) or evolutionary computation are examples of sub-symbolic approaches in ML. Interestingly so far, none of the actual methods available seem to promise a path to AGI.

Yet, despite having at least some general definitions at hand, the common discussion usually ignores those and therefore the range of AI-assigned functionality reaches from applying traditional statistics to using machine learning (ML) techniques up to solely movie inspired ideas or even generally to 'highly complex information systems', as in the official 'Social Principles of Human-centric AI' of Japan (Council for Social Principles of Human-centric AI 2019).

In the following, we will concentrate on artificial neural networks to illustrate the fallacies and pitfalls of questionably used language. The focus on ANN in this text is in line with the current debate of AI, where AI is predominantly used synonymously with machine learning using artificial neural networks (Eberl 2018). Nevertheless, the problems mentioned here also apply to debates concerning other forms of AI, when a similar terminology is being used.

Key drivers for the current AI renaissance are the successes of applying artificial neural networks to huge amounts of data now being available and using new powerful hardware. Although the theoretical foundations of the concepts used were conceived as early as the 1980s, the performance of such a system has improved to such an extent over the last years, that they can now be put to practical use in many new use cases, sometimes even in real-time applications such as image or speech recognition. Especially if huge data sets for training are available, depending on the task results can be much better than traditional symbolic approaches where information is written into databases for explicit knowledge representation.

Before we analyse the language being used to describe the functionality, we should have a look at the inner workings of artificial neural networks to have a base for scrutinising terminologies.

### Basic Structure of Artificial Neural Networks

Artificial neural networks are an approach of computer science to solve complex problems that are hard to explicitly formulate, or more concretely: to program. Those networks are inspired by the function of the human brain and its network of neurons; however, the model of a neuron being used is very simplistic. Many details of biological neuronal networks, such as myelination or ageing (Hartline 2009), are left out, as well as new mechanisms, such as backpropagation (Crick 1989), are introduced. Trying to follow the original model, each artificial neuron, the smallest unit of such systems, has several inputs and one output. In each artificial neuron, the inputs are weighted according to its configuration and then summed up. If the result exceeds a certain defined threshold the neuron is triggered, and a signal is passed on to the output. These neurons are usually formed into 'layers', where each layer's outputs are the next layer's inputs. The resulting artificial neural network thus has as its input the individual inputs of the first layer, and as its output the individual outputs of the last layer. The layers in between are usually called 'hidden' layers and with many hidden layers an artificial neural network is usually called 'deep'.

In the practical example of image recognition, the input would consist of the colour values of all distinct pixels in a given image and the output would be the probability distribution among the predefined set of objects to recognise.

### Configuring the Networks

From a computer science point of view ANNs are very simple algorithms, since the signal paths through the connections of the network can easily be calculated by mathematical equations. After all, it is the variables of this equation (weights, thresholds, etc.) that accord the powerful functionality to ANNs. So ANNs are basically simple programs with a very complex configuration file and there are various ways of configuring artificial neural networks, which will now briefly be described. Building such a network involves certain degrees of freedom and hence decisions, such as the number of artificial neurons, the number of layers, the number and weights of connections between artificial neurons and the specific function determining the trigger behaviour of each artificial neuron. To properly recognise certain patterns in the given data – objects, clusters etc. – all those parameters need to be adjusted to a use case. Usually there are best practices how to initially set it up; then the artificial network has to be further improved step by step. During this process the weights of the connections will be adjusted slightly in each step, until the desired outcome is created, may it be the satisfactory detection of cats in pictures or the clustering of vast data in a useful way. Those training cycles are often done with a lot of labelled data and then repeated until the weights do not change any more. Now it is a configured artificial neural network for the given task in the given domain.

*Speaking about the Networks*

Now we will take a closer look at how computer scientists speak about this technology in papers and in public, and how those utterances are carried into journalism and furthermore into politics. As mentioned above, the description of ANNs as being inspired by the human brain already implies an analogy which must be critically reflected upon. Commonly used ANNs are usually comparatively simple, both in terms of how the biochemical properties of neurons are modelled but also in the complexity of the networks themselves. A comparison: the human brain consists of some 100 billion neurons while each is connected to 7,000 other neurons on average. ANNs on the other hand are in the magnitude of hundreds or thousands of neurons while each is connected to tens or hundreds of other neurons. This difference in orders of magnitude entails a huge difference in functionality, let alone understanding them as models of the human brain. Even if to this point the difference might only be a matter of scale and complexity, not principle, we have no indication of that changing anytime soon. Thus, using the notion of 'human cognition' to describe ANN is not only radically oversimplifying, it also opens up the metaphor space to other neighbouring yet misleading concepts. For example, scientists usually do not speak of networks being configured but being 'trained' or doing '(deep) learning'. Along those lines are notions like 'recognition', 'acting', 'discrimination', 'communication', 'memory', 'understanding' and, of course, 'intelligence'.

*Considering Human Related Concepts*

When we usually speak of 'learning', it is being used as a cognitive and social concept describing humans (or, to be inclusive, intelligent species in general) gaining knowledge as individual learner or as a group, involving other peers, motivations, intentions, teachers or coaches and a cultural background (Bieri 2017). This concept includes the context and a whole range of learning processes being researched, tested and applied in the academic and practical fields of psychology of learning, pedagogy, educational science (Piaget 1944) and neuroscience (Kandel and Hawkins 1995). This is a substantial difference to the manual or automated configuration of an ANN using test sets of data. Seen in this light, the common notion of 'self-learning systems' sounds even more misplaced. This difference in understanding has great implications, since, for example, an ANN would never get bored with its training data and therefore decide to learn something else or simply refuse to cooperate (Weizenbaum 1976); metaphors matter, no Terminator from the movies in sight.

'Recognition' or 'memory' are also very complex concepts in the human realm. Recognising objects or faces requires attention, focus, context and – depending on one's school of thought – even consciousness or emotions. Human recognition is therefore completely different from automatically finding differences of brightness in pictures to determine the shape and class of an expected object (Goodman 1976). Furthermore, consciously remembering something is a highly complex process for humans which is more comparable to living through imagined events again and by that even changing what is being remembered. Human memory is therefore a very lively and dynamic process, and not at all comparable to retrieving accurate copies of stored data bits (Kandel and Hawkins 1995).

Especially the notions of 'action' or even 'agency' are highly problematic when being applied to computers or robots. The move of a computer-controlled robotic arm in a factory should not be called a robot's 'action', just because it would be an 'action' if the arm belonged to a human being. Concerning human actions, very broad and long-lasting discussions at least in philosophy and the social sciences already exist, note the difference between 'behaviour' and 'action' (Searle 1992). The former only focuses on observable movement, whereas the latter also includes questions of intention, meaning, consciousness, teleology, world modelling, emotions, context, culture and much more (Weizenbaum 1976). While a robot or a robotic arm can be described in terms of behavioural observations, its movements should not easily be called actions (Fischer and Ravizza 1998).

Similarly complex is the notion of 'communication' in a human context, since communication surely differs from simply uttering sounds or writing shapes. 'Communication' requires a communication partner, who knows that the symbols used have been chosen explicitly with the understanding that they will be interpreted as deliberate utterances (von Savigny 1983). Communication therefore needs at least the common acknowledgement of the communicational process by the involved parties, in other words an understanding of each other as communicating (Watzlawick 1964). A 'successful' communication is then the result of both parties agreeing that it was successful and therefore the creation of a common understanding. Hence, the sound of a loudspeaker or the text on a screen does not constitute a process of communication in the human sense, even if their consequences are the production of information within the receiving human being. If there is no reflection of the communication partner, no deliberation, no freedom of which symbols to choose and what to communicate one should not easily apply such complex notions as 'communication' outside its scope without explanation.

Furthermore, the concept of 'autonomy' – as opposed to heteronomy or being externally controlled – is widely used nowadays when dealing with artificial intelligence, may it be concerning 'intelligent' cars or 'advanced' weapon systems. Although starting in the last century even human autonomy has been largely criticised within the social sciences (some even say completely deconstructed, Krähnke 2006) since individuals are largely influenced by culture, societal norms and the like, the concept of autonomy seems to gain new traction in the context of computer science. Yet, it is a very simplistic understanding of the original concept (Gerhardt 2002). Systems claimed to be 'autonomous' heavily depend on many factors, e.g., a stable, calculable environment, but also on programming, tuning, training, repairing, refuelling and debugging, which are still traditionally done by humans, often with the help of other technical systems. In effect those systems act according to inputs and surroundings, but they do not 'decide' on something (Kreowski 2018), certainly not as humans do (Bieri 2001). Here again, the system can in principle not contemplate its actions and finally reach the conclusion to stop operating or change its programmed objectives autonomously. Hence, artificial intelligence systems – with or without ANN – might be highly complex systems, but they are neither autonomous nor should responsibility or accountability be attributed to them (Fischer and Ravizza 1998). Here we see one concrete instance of the importance of differentiating between the domain-specific ANI and universal AGI (Rispens 2005). This clarification is not meant to diminish the technical work of all engineers involved in such 'autonomous' systems, it is purely a critique about how to adequately contextualise and talk about such systems and its capabilities in non-expert contexts.

## Instances and Consequences

After having briefly touched upon some areas of wrongly used concepts, we can take a look at concrete examples, where such language use specifically matters.

A very interesting and at that time widely discussed example was Google's 'Deep Dream' image recognition and classification software from 2015, codename 'Inception'. As described above, ANNs do not contain any kind of explicit models; they implicitly have the 'trained' properties distributed within their structures. Some of those structures can be visualised by inserting random data – called 'noise' – instead of actual pictures. In this noise, the ANN then detects patterns exposing its own inner structure. What is interesting are not the results – predominantly psychedelic imagery – but the terminology being used in Google's descriptions and journalists' reports. The name 'Deep Dream' alone is already significant, but also the descriptive phrases 'Inside an artificial brain' and 'Inceptionism' (Mordvintsev and Tyka 2015). Both phrases (deliberately) give free rein to one's imagination. In additional texts provided by Google, wordings such as 'the network makes decisions' accumulate. Further claims are that it 'searches' for the right qualities in pictures, it 'learns like a child' or it even 'interprets' pictures. Using this misleading vocabulary to describe ANNs and similar technical artefacts, one can easily start to hope that they will be able to learn something about the fundamentals of human thinking. Presumably those texts and descriptions have been written for the primary purpose of marketing or public relations, since they explain little but signify the abilities and knowledge of the makers, yet that does not diminish the effect of the language used. For many journalists and executive summary writers or even the interested public those texts are the main source of information, not the hopefully neutral scientific papers. In effect, many of those misleading terms where widely used, expanded on and by that spread right into politician's daily briefings, think-tank working papers and dozens of management magazines, where the readers are usually not aware of the initial meanings. This distorted 'knowledge' then becomes the basis for impactful political, societal and managerial decisions.

Other instances where using wrong concepts and wordings mattered greatly are in car crashes involving automated vehicles, e.g., from companies like Uber, Google or Tesla. For example, in 2018 a Tesla vehicle drove into a parked police car in California, because the driver had activated the 'autopilot' feature and did not pay attention to the road. This crash severely exposed the misnomer. The driver could have read the detailed 'autopilot' manual before invoking such a potentially dangerous feature, yet, if this mode of driving had been called 'assisted driving' instead of 'autopilot', very few people would have expected the car to autonomously drive 'by itself'. So, thinking about a car having an autopilot is quite different from thinking about a car having a functionality its makers call 'autopilot'. Actually reading into Tesla's manuals, different levels of driving assistance are being worked on, e.g., 'Enhanced Autopilot' or 'Full Self-Driving', whereas the latter has not been implemented so far. Further dissecting the existing 'autopilot' feature one finds it comprises different sub-functionalities such as Lane Assist, Collision Avoidance Assist, Speed Assist, Auto High Beam, Traffic Aware Cruise Control or Assisted Lane Changes. This collection of assistance technologies sounds very helpful, yet it does not seem to add up to the proclaimed new level of autonomous driving systems with an autopilot being able to 'independently' drive by itself.

Those examples clearly show how a distinct reality is created by talking about technology in certain terms, yet avoiding others. Choosing the right terms, is not always a matter of life and death, but they certainly pre-structure social and societal negotiations regarding the use of technology.

*Malicious Metaphors and Transhumanism*

Suddenly we arrive in a situation where metaphors are not only better or worse for explaining specifics of technology, but where specific metaphors are deliberately being used to push certain agendas; in Tesla's case to push a commercial and futurist agenda. Commercial because of using 'autonomy' as a unique selling point for cars and futuristic, as it implies that 'autonomy' is a necessary and objective improvement for everyone's life and the society as a whole. Generally, most innovative products involving 'artificial intelligence' and 'next generation technology' are being communicated as making 'the world a better place', 'humans more empowered' or 'societies more free' by the PR departments of the offering companies and spread even further by willing believers and reporting journalists. The long-standing effects of metaphors let loose can also be seen vividly in the discourse about transhumanism, where humans themselves, even humankind as a whole, should be enhanced and improved using (information) technology, predominantly by using AI. Here again the proponents either really believe in or profit from those narratives, or both (Kurzweil 2005).

In this discourse all mistakes of the AI terminology can be observed fully developed with many consequences, since when we pose the transhumanist question regarding how information technology can help human beings the answer is usually 'enhancement'. Yet the notion of 'enhancement' is being used in a very technical way, ignoring its fundamental multiplicity of meanings. With information technology, so the argument from the classic flavour of transhumanism goes, we will soon be able to fix and update the human operating system: merging with intelligent technical systems will make our brain remember more faces, forget less details, think faster, jump higher, live longer, see more sharply, be awake longer, be stronger, hear more frequencies and even create new senses – exactly how a technologist would imagine what new technology could deliver for humanity (Kurzweil 2005). More recent concepts see humans and AI systems in a cooperative even symbiotic relationship. Those concepts exemplify the direction of imagination once we assume there are truly 'intelligent' systems with 'agency' who can 'decide' and 'act'.

However interestingly the underlying and implicit assumption is a very specific – to be precise: technical – understanding of what is considered 'good' or 'desirable'. But does every human or even the majority primarily want to remember more, forget less, live longer or run faster? Are those aspects even the most pressing issues we want technology to solve? In addition, not only do those fantasies happily follow along the lines of the neo-liberal logic of applying quantification, competition, performance and efficiency into all aspects of life, they also unconsciously mix in masculinist – even militarist – fantasies of power, control, strength and subjugation of the natural or finally correcting the assumed defective (Schmitz and Schinzel 2004).

As valid as those opinions concerning optimisations are, still it is important that views like that imply absolute values and are incompatible with views which put social negotiation, non-mechanistic cultural dynamics or in general pluralistic approaches in their centre. To structure the discourse, I call those conflicting groups of views *regimes of enhancement*. Clearly it is not possible to 'enhance' a human being with technically actualised immortality, if this person does not want to live forever or does not find it particularly relevant. Many other conflicting views can be thought of. However, the mere acceptance of the concept of *regimes* already breaks any claim for absoluteness and opens the door for discussing different understandings of 'enhancements'. Accepting this already makes any positions somehow compatible and allows for individual or even societal endeavours of creative re-interpretations of the concept of transhumanism itself (Haraway 1991).

The transhumanist discourse outlines the consequences of not reflecting on core notions like 'enhancement' in the same way as it is consequential not to reflect on 'intelligence' in the AI discourse (Bonsiepen 1994). Broken visions and faulty applications are to be expected. Furthermore, this kind of language also shapes and attracts a certain kind of mindset where the above mentioned reductionist metaphors are not even used as metaphors anymore, but as accurate descriptions of the world (Coy 1993).

## Constructive Wording

So, next time decision-makers and journalists will be asked about possibilities of technology they will surely remember having heard and read about computers winning Chess and Go, driving cars, recognising speech, translating text, managing traffic and generally finding optimal solutions to given problems (Dreyfus 1972). But using deficient anthropomorphisms like 'self-learning', 'autonomous' or 'intelligent' to describe the technical options of solving problems will lead to malicious decisions (The Royal Society 2018, 7–8).

Surely the best solution for this problem would be to completely change the terminology, but since large parts of the above mentioned are fixed scientific terms, a clean slate approach seems unrealistic. Therefore, at least in interdisciplinary work, science journalism activities or political hearings, a focus should be put on choosing the appropriate wording by scientists and (science) journalists. Only then policy and decision-makers have a chance to meaningfully grasp the consequences of their actions. In addition, interdisciplinary research could also get a more solid (communication) ground. Of course, this change of terms will not be the end of discipline-limited jargon in AI, but it would surely increase the efficiency of exchange between the different fields.

For concretely deciding which terms to use and which words to change it would be generally preferable to have some kind of criteria. Following the above descriptions, the used terminology should be as close as possible to the technical actualities while at the same time avoiding:

- technical terms that have a connotation in common language reaching far beyond the actual technical function, e.g., recognition, agent, communication, language, memory, training, senses, etc. since they will be understood as metaphors

- anthropomorphisms which are not technical terms but usually used as metaphors to describe technical details, e.g., thinking, (re)acting, deciding, remembering, etc.
- concepts widely used in popular science, media and science fiction implying a completely different meaning e.g., intelligent machine, android, self-improving, autopilot, etc.

Certainly those words can be replaced by more fitting vocabulary. Depending on context 'remembering' could be paraphrased by 'implicitly stored in configuration', 'learning' by 'changing/improving configuration', 'recognition' by 'detection', 'intelligent' by 'automated' (cf. Butollo 2018), 'action' by 'movement' or 'response', 'decision' or 'judgement' by 'calculation' (cf. Weizenbaum 1976), and 'communication' by 'indication' or 'signalling'. However, terms like 'agency' and 'autonomy' should be discarded entirely, since they are neither accurate or necessary nor helpful; they are just completely misleading.

Being aware that this change might also bear consequences for scientific grant proposals which usually have to sound societally important, scientifically innovative and relevant, it is imperative here too as part of science ethics to reflect on the wider consequences of the language used to communicate. Admittedly it should be noted that those suggestions won't be applied by speakers who are deeply convinced of such metaphors fitting the subject matter, yet, they would then be clearly visible as such.

## Closing Remarks

Technology is used and politically decided upon perceived functionality, not upon the actually implemented functionality. However, communicating functionality is much more driven by interests than creating the actual technology. Therefore, attribution ascription is a very delicate and consequential issue that paints a differentiated picture of the consequences of careless use of terms. If relevant decision-makers in politics and society are (really) convinced at some point that these 'new' artificial neural networks can develop an understanding of things or properly interpret facts, nothing would stand in the way of their use for socially or politically sensitive tasks like deciding about social benefits, teaching children or judging court cases. Here the difference between 'judging' and judging, 'acting' and acting play out. If one acts in the social science meaning of the word, one has to take responsibility for one's actions, if a computer only 'acts', used as a metaphor, responsibility is blurred.

Hence, especially computer professionals but also scientific journalists should follow the professional responsibility to be more sensitive about the criticised misleading metaphors and in effect to change them to more fitting ones. The danger here does not lie in incompletely understanding computers or AI but in not understanding them while thinking that they have been understood. A possible way out of this tricky situation is certainly more disciplinary openness towards interdisciplinary research and communication. Especially the discipline of computer science could embrace this kind of exchange much more, from student curricula to research projects. This would maybe not so much change their disciplinary core work but it would contextualise this work, create better accessibility for other less technical fields and produce overall more useful results. Naturally, this would require all parties involved to speak to each other but also to listen and teach each other ways of looking at the world. Of course, and not least those ventures must be encouraged and facilitated by field leaders, research grant givers and research politics alike.

So, if technological discussions and societal reflections on the use of technology are to be fruitful, scientists and (science) journalists alike have to stop joining the buzzword-driven language game of commercial actors and AI believers alike, which does neither help with solutions nor advances science. It merely entertains our wishful thinking of how magical technology should shape the future. Finally, we record that a chess computer will never get up and change its profession, exponential growth in computing power does so far not entail more than linear growth of cognitive-like functionality, and the fear of computers eliminating all human jobs is a myth capable of inciting fear since at least 1972.

But maybe, indeed, any sufficiently advanced technology is indistinguishable from magic – to the layperson – but we also have to conclude that this 'magic' is being constructed and used by certain expert 'magicians' to advance their own interests and agendas, or that of their masters (Hermann 2020). So not even such magical interpretation would spare us the necessity to pay attention to power, details and debate (Kitchin 2017). This chapter tries to constructively be a part of this interdisciplinary project.

## References

Bieri, P. 2001. *Das Handwerk der Freiheit*. Munich: Hanser.
Bieri, P. 2017. *Wie wäre es, gebildet zu sein?* Munich: Komplett Media GmbH.
Bonsiepen, L. 1994. Folgen des Marginalen. Zur Technikfolgenabschätzung der KI. In: G. Cyranek and W. Coy (Eds.), *Die maschinelle Kunst des Denkens. Theorie der Informatik*. Braunschweig/Wiesbaden: Vieweg.
Butollo, F. 2018. Automatisierungsdividende und gesellschaftliche Teilhabe. *Regierungsforschung.de*, NRW School of Governance. Retrieved from https://regierungsforschung.de/wp-content/uploads/2018/05/23052018_regierungsforschung.de_Butollo_Automatisierungsdividende.pdf
Council for Social Principles of Human-centric AI. 2019. *Social Principles of Human-Centric AI*. Council for Social Principles of Human-centric AI: Japan.
Coy, W. 1992. Für eine Theorie der Informatik! In: W. Coy et al. (Eds.), *Sichtweisen der Informatik. Theorie der Informatik*, pp. 17–32. Wiesbaden: Vieweg +Teubner Verlag.

Coy, W. 1993. Reduziertes Denken. Informatik in der Tradition des formalistischen Forschungsprogramms. *Informatik und Philosophie* 22, 31–52.

Crick, F. 1989. The Recent Excitement About Neural Networks. *Nature* 337 (6203), 129–132.

Dreyfus, H. L. 1972. *What Computers Can't Do*. New York: Harper & Row.

Eberl, U. 2018. Was ist Künstliche Intelligenz – was kann sie leisten? *Aus Politik und Zeitgeschichte,* 6–8 (2018), 8–14.

Fischer, J. M. and Ravizza, M. 1998. *Responsibility and Control: A Theory of Moral Responsibility*. Cambridge: Cambridge University Press.

Gerhardt, V. 2002: *Freiheit als Selbstbestimmung*. In: Wobus A. N. et al, (Eds.) *Nova Acta Leopoldina*, Issue 324, Volume 86, Deutsche Akademie der Naturforscher, Leopoldina, Halle (Saale).

Gevarter, W. B. 1985. *Intelligent Machines: Introductory Perspective on Artificial Intelligence and Robotics*. Englewood Cliffs, NJ: Prentice Hall.

Goodman, N. 1976. *Languages of Art: An Approach to a Theory of Symbols*. Indianapolis, MA: Hackett Publishing.

Haraway, D. 1991. A Cyborg Manifesto: Science, Technology, and Socialist-Feminism in the Late Twentieth Century. *Simians, Cyborgs and Women: The Reinvention of Nature*, pp. 149–181. New York: Routledge.

Hartline, D. K. 2009. *What is Myelin?* Cambridge: Cambridge University Press.

Hermann, I. 2020. Künstliche Intelligenz in der Science-Fiction: Mehr Magie als Technik. *Von Menschen und Maschinen: Interdisziplinäre Perspektiven auf das Verhältnis von Gesellschaft und Technik in Vergangenheit, Gegenwart und Zukunft. Proceedings der 3. Tagung des Nachwuchsnetzwerks 'INSIST', 5–7 October 2018, Karlsruhe (INSIST-Proceedings 3)*.

Kandel, E. R. and Hawkins, R. D. 1995. Neuronal Plasticity and Learning. In: R. D. Broadwell (Ed.), *Neuroscience, Memory, and Language. Decade of the Brain, Vol. 1*, S. 45–58. Library of Congress: Washington, DC.

Kitchin, R. 2017. Thinking Critically About and Researching Algorithms. *Information, Communication & Society* 20 (1), 14–29.

Krähnke, U. 2006. Selbstbestimmung. *Zur gesellschaftlichen Konstruktion einer normativen Leitidee*. Weilerswist: Velbrück Verlag.

Kreowski, H.-J. 2018. Autonomie in Technischen Systemen. *Leibniz Online* 32.

Kurzweil, R. 2005. *The Singularity Is Near: When Humans Transcend Biology*. London: Penguin.

Mitchell, T. 1997. *Machine Learning*. New York: McGraw Hill.

Mordvintsev, A. and Tyka, M. 2015. Inceptionism: Going Deeper into Neural Networks. *Google AI Blog*. 17 June 2015. Retrieved from: https://ai.googleblog.com/2015/06/inceptionism-going-deeper-into-neural.html.

Morozov, E. 2013. *To Save Everything, Click Here: The Folly of Technological Solutionism*. New York: PublicAffairs.

Mosco, V. 2004. *The Digital Sublime: Myth, Power, and Cyberspace*. Cambridge, MA: MIT Press.

Passig, K. and Scholz, A. 2015. *Schlamm und Brei und Bits – Warum es die Digitalisierung nicht gibt*. Stuttgart: Klett-Cotta Verlag.

Piaget, J. 1944. *Die geistige Entwicklung des Kindes*. Zurich: M.S. Metz.

Rispens, S. I. 2005. *Machine Reason: A History of Clocks, Computers and Consciousness*. Doctoral Thesis, University of Groningen.

Royal Society, The. 2018. *Portrayals and Perceptions of AI and Why They Matter*. London. Retrieved from: http://lcfi.ac.uk/media/uploads/files/AI_Narratives_Report.pdf.

von Savigny, E. 1983. Zum Begriff der Sprache – Konvention, Bedeutung, Zeichen, Stuttgart: Reclam.

Schmitz, S. and Schinzel, B. 2004. *Grenzgänge: Genderforschung in Informatik und Naturwissenschaften*. Ulrike Helmer Verlag.

Searle, J. R. 1992. *The Rediscovery of Mind*, Cambridge, MA: MIT Press.

Turing, A. 1950. Computing Machinery and Intelligence. *Mind* LIX (236), 433–460. DOI: https://doi.org/10.1093/mind/LIX.236.433.

Watzlawick, P. 1964. *An Anthology of Human Communication*. Palo Alto, CA: Science and Behavior Books.

Weizenbaum, J. 1976. *Computer Power and Human Reason: From Judgment to Calculation*. San Francisco: W. H. Freeman.